\documentclass[a4paper]{jpconf}
\usepackage{graphicx}
\begin{document}

\title{The Oblique Basis Method from an Engineering Point of View?}

\author{V. G. Gueorguiev}

\address{California State University Stanislaus, Turlock, CA 95382}

\ead{vesselin@mailaps.org}

\begin{quotation}
\small \em "Take the world from another point of view." - R. Feynman
\end{quotation}

\begin{abstract}
The oblique basis method is reviewed from engineering point of view related to vibration and control theory. Examples are used to demonstrate and relate the oblique basis in nuclear physics to the equivalent mathematical  problems in vibration theory. The mathematical techniques, such as principal coordinates and root locus, used by vibration and control theory engineers are shown to be relevant to the Richardson - Gaudin pairing-like problems in nuclear physics.
\end{abstract}

\section{Introduction}

It is well known and commonly experienced observation that ``... mathematical concepts turn up in entirely unexpected connections ..." \cite{1960}. The surprise of finding a mathematical structure applicable to seemingly unrelated fields of study is somehow proportional to our ignorance of one or the other subject, even though we may have studied the underlining mathematical field previously. The following contribution attests to this situation by drawing the analogies and pointing examples from my research topic in nuclear physics and a more recent engineering course in vibration and control theory which I had the pleasure to teach. This reinforces my view that one should keep an open mind and must often try to ``Take the world from another point of view" \cite{1973}. In this respect, this article is devoted to the discussion of two major applications of a linear algebra: matrix diagonalization and location of zeros of complex algebraic systems. The first application is in the subject of complex many-body systems like nuclei \cite{2001} while the second is their application in the filed of vibration and control theory \cite{2011}.  Although presenting the classical mechanics problem followed by its quantum mechanics framework within which the mathematical tools reside may be the better pedagogical point of view, we will use the opposite order of the exposition, due to the fact that the audience to which the material was presented is more familiar with my nuclear physics research framework.

In what follows, the article justifies the oblique basis approach to complex problems and introduces the two important collective models commonly used in nuclear physics. Then the oblique basis method is stated as it is pertained to nuclear physics. The engineering problem is stated next and the relevant mathematical techniques, such as Evans root-locus method \cite{1948} for studying the stability of a system, are outlined. Finally, we go back to the nuclear physics subject and show the relevance of the  Evans root-locus method for the study of exactly solvable nuclear pairing interaction.

Before continuing the article with its main subject, I would like to acknowledge that, even though the oblique basis idea has  been developed mostly as a collaborative effort \cite{2001, 2005, 2007}, in my view it has been stimulated and supported strongly by my Ph.D. adviser Professor Jerry Draayer, in whose honor this conference has been held.
\section{Important Exactly Solvable  2-body Nuclear Interactions}

Exactly solvable problems such as harmonic oscillator and other group-symmetry based models in physics have been essential in the qualitative and quantitative understanding of physics phenomena. The need of better, more accurate, and easily replicable quantitative results has naturally introduced the use of powerful computers to the physics research. Never the less, exactly solvable problems are still essential to our understanding of nuclear structure \cite{2010} and reactions \cite{2007}, especially in the limit of weak perturbation to the exact limit \cite{2005}. In particular, the nuclear shell-model in the limit of independent single particle motion is exactly solvable; however, when a two-body perturbation interaction is added then it becomes the cornerstone of the configuration mixing approach to the corresponding many-body problem \cite{2010}. When the interaction between the nucleons becomes significant the perturbation method is no-longer applicable and cannot reproduce the correct behavior of the system. Examples of such cases are well-deformed nuclei where the rotational structure of the spectra becomes more prominent or near-closed-shell nuclei where pairing phenomenon take over the low-energy spectra of the nuclei.

By using Fock space techniques within the independent particle model with traditional single-particle level labels, one can consider two important collective interactions in nuclei. These interactions represent two well known exactly solvable systems:
\begin{itemize}
\item the Elliott's SU(3) model with degenerate single-particle harmonic oscillator energies 
($\varepsilon _{i}=\varepsilon _{j}= \varepsilon$) and quadrupole-quadrupole interaction $Q\cdot Q$ \cite{2010}: 
$$ H=\varepsilon N - \frac{\chi}{2} Q\cdot Q,$$
\item the nuclear pairing ($P^{+}P$) interaction with non-degenerate single-particle energies
($\varepsilon _{i}$):
$$H=\varepsilon _{i}N_{i}+GP^{+}P. $$
\end{itemize}

\section{T=1 Proton-Neutron Pairing - SO(5) RG-model}
While the exact solvability of the  Elliott's SU(3) model is more or less well understood from group-theoretical point of view, the exact solvability of the particle number conserving pairing interaction has much less known implementations, even though the solution has been published since 1962 by Richardson \cite{1962}. The exploration and implementation of the Richardson-Gaudin particle number conserving pairing interaction may have been hindered by its more complicated algebraic structure as seen in the work of Gaudin \cite{1976} and Ushveridize \cite{1994}. However, the advancement of the computer power and recent understanding of the relevant mathematical and computational structure has brought a renewed interest in the topic \cite{1997, 2002} along with the understanding of the integrability of the pairing interaction \cite{2002Links, 2002Asorey}. The pairing has even been extended to an A-body generalized pairing interaction \cite{2003}. Finally, the full T=1 proton-neutron pairing has been formulated within the frame work of the Richardson-Gaudin algebras \cite{2006}.

Here we briefly summarize the main results on the Richardson-Gaudin pairing with the purpose of pointing out some of the possible practical implementation issues. We recall the 10 generators of the SO(5) algebra in
a representation suited for nuclear physics problems \cite{2006}. The three $T=1$ pair-creation operators are defined as: 
$\hat b^\dag_{-1,i}=\hat n_i^\dag\hat n_{\bar\imath}^\dag$, $\hat
b^\dag_{0,i}=(\hat n_i^\dag\hat p_{\bar\imath}^\dag+\hat
p_i^\dag\hat n_{\bar\imath}^\dag)/\sqrt{2}$, and $\hat
b^\dag_{+1,i}=\hat p_i^\dag\hat p_{\bar\imath}^\dag$, where $n$
and $p$ refer to neutrons and protons, respectively, and $i$
labels a single-particle basis (with ${\bar\imath}$ its
time-reversed state) that may be associated with the spherical
shell-model basis $i\equiv jm$ or with an axially-symmetric
deformed basis $i\equiv\alpha m$. The three pair-annihilation
operators are $\hat b_{-1,i}$, $\hat b_{0,i}$, and $\hat
b_{+1,i}$. The three components of the isospin operator [$\hat
T_{+,i}=(\hat p_i^\dag\hat n_i+\hat p_{\bar\imath}^\dag\hat
n_{\bar\imath})/\sqrt{2}$, $\hat T_{0,i}=(\hat p_i^\dag\hat
p_i+\hat p_{\bar\imath}^\dag\hat p_{\bar\imath})/2 -(\hat
n_i^\dag\hat n_i+\hat n_{\bar\imath}^\dag\hat n_{\bar\imath})/2$,
and $\hat T_{-,i} =(\hat n_i^\dag\hat p_i+\hat
n_{\bar\imath}^\dag\hat p_{\bar\imath})/\sqrt{2}$] close the ${\rm
SU}_T(2)$ subalgebra of SO(5). These 9 operators together with the
number operator $\hat N_{i}=\hat p_i^\dag\hat p_i+\hat
p_{\bar\imath}^\dag\hat p_{\bar\imath}+ \hat n_i^\dag\hat n_i+\hat
n_{\bar\imath}^\dag\hat n_{\bar\imath}$ define the SO(5) algebra.
The pn-pairing Hamiltonian with a pairing strength $g$ and isospin breaking $\Delta$ is given by the expression \cite{2006}:
\begin{eqnarray}
\hat H&=&\sum_j\varepsilon_j
\left(\hat N_j+\Delta\hat N_{p,j}\right)
+\frac{g}{2}\hat T\cdot\hat T+
g\!\!\!\!\sum_{\mu jmj^\prime m^\prime}\!\!\!
\hat b_{\mu,jm}^\dag\hat b_{\mu,j^{\prime }m^{\prime }},
\label{Ham1}
\end{eqnarray}
with the eigenvalues of~(\ref{Ham1}) given by the expression:
\begin{eqnarray}
E=\sum_{\alpha=1}^M e_\alpha+
\frac{\Delta}{2}\!\sum_{\beta=1}^{M+T_0+t}\!w_\beta+
\sum_j\varepsilon_j \left[\frac{v_j}{2}(2+\Delta)-\Delta
t_j\right] + \frac{g}{2} T_0(T_0-1),\label{ener}
\end{eqnarray}
where $v_i$ is the seniority of each
$i$ level, {\it i.e.} the number of fermions not paired in
time-reversed states with isospin $T=1$, $t_i$ is the isospin of
the unpaired fermions, $t=\sum_it_i$, $M$ is the number of $T=1$
time-reversed pairs, and $T_0$ is the $z$-axes component of the total
isospin, {\it i.e.} the eigenvalue of the operator $\hat
T_0=\sum_i\hat T_{0,i}$. The total number of nucleons is
$N=N_p+N_n=2M+\sum_i v_i$ whereas their difference is
$N_p-N_n=2T_0$. The quantum numbers $M$, $T_0$, $v_i$, and $t_i$
are conserved; $T$ is also conserved if $\Delta=0$.

The variables  $e_\alpha$ and $\omega_\beta$ are solutions of the corresponding Richardson-Gaudin equations \cite{2006}:
\begin{eqnarray}
\frac{1}{g}&=&
\frac{1}{2}\sum_{i=1}^L
\frac{v_{i}+2t_i-2}{z_i-e_\alpha}+
\sum_{\alpha^\prime(\neq\alpha)=1}^M
\frac{2}{e_{\alpha^\prime}-e_\alpha}-\sum_{\beta=1}^{M+T_0+t}
\frac{1}{\omega_\beta-e_\alpha},
\label{Richeqs}\\
\frac{\Delta }{g}&=&
\sum_{\beta^\prime(\neq\beta)=1}^{M+T_0+t}
\frac{2}{\omega_{\beta^\prime}-\omega_\beta}-
\sum_{\alpha=1}^M
\frac{2}{e_\alpha-\omega_\beta}-
\sum_{i=1}^L\frac{2t_i}{z_i-\omega_\beta}.
\label{pnRicheqs}
\end{eqnarray}
where variables $z_j$ are related to the single particle energies $\varepsilon_j$ ($z_j=2 \varepsilon_j$ ) and the indexes are specialized to a spherical basis $i \equiv jm$. 

Thus, the problem of solving the pairing hamiltonian is reduced to solving the above non-linear equations for $e_\alpha$ and $\omega_\beta$. In practical terms, without computers in the old days, this would mean hard work by hand. Nowadays, there are sufficient computer power to solve these equations by starting at a small values of coupling constants $g$ and $\Delta$ and gradually increasing these values to the desired magnitudes. However, there is an apparent singularity problem when some of the pair energies become close to the a single particle energy level, that is, when some of the denominators get close to zero. This problem can be resolved by trial and error search on the values pass the singularity point - this is similar to the problem of finding initial configurations for the values of $e_\alpha$ and $\omega_\beta$ to start the process with. Alternatively, one can use small complex component of the single particle energies or the Evans root-locus method to be touched upon in the last section.

\section{Oblique Basis in Nuclear Physics}
In second quantized form the effective Hamiltonian for a nuclear system is:
\begin{equation}
H=\sum_{i}\varepsilon _{i}a_{i}^{+}a_{i}+\frac{1}{4}
\sum_{i,j,k,l}V_{ij,kl}a_{i}^{+}a_{j}^{+}a_{k}a_{l}. \label{H=aa+aaaa}
\end{equation}
Here, $\varepsilon _{i}$ are single-particle energies derived from
excitation spectra of one valence particle system and $V_{kl,ij}$ are two-body matrix elements viewed as a residual interaction beyond the single-particle mean-field confining potential. 

When the residual interaction can be treated as perturbation then one can apply perturbation theory using the basis of the exactly solvable limit. If there are two or more exactly solvable limits then there are naturally two or more possible perturbation treatments that are based on their corresponding basis sets. However, if the strengths of the interactions are such that perturbation theory is not applicable, that is, there is a strong mixing, then a combination of relevant basis sets can be used to approach the problem. Thus, the oblique basis consists of two or more basis sets. In this sense the oblique basis idea has been applied to the case when the Elliott's SU(3) is broken due to the non-degenerate single-particle energies \cite{2001} and also illustrated on simple toy model of a particle in a box \cite{2005}.

In our simple example  \cite{2001} the two basis sets were: spherical shell-model states (ssm-states) expressed in spherical single-particle coordinates ($nlj$) that are eigenvectors of the single particle non-interaction hamiltonian, while the second set has a good SU(3) structure (su3-states) that track nuclear deformation. The SU(3) basis set is given in cylindrical single-particle coordinates. Such oblique calculations result in a nonorthogonal oblique basis and require an evaluation of the matrix elements of physical operators plus a knowledge of the scalar product ($e_{\alpha}\cdot E_{i}$) related to the overlap matrix. Schematically, these basis vectors are labeled by $\alpha = 1$,..., dim(ssm-basis) and $i = 1$,..., dim(su3-basis) and the corresponding matrixes can be represented in the following way:

\begin{eqnarray}
\mathrm{basis\quad vectors} &:&\mathrm{\quad}\left(\begin{array}{l}
e_{\alpha} :\mathrm{ssm\ - \ basis} \\
E_{i} :\mathrm{su3\ - \ basis}
\end{array}
\right), \label{Basis vectors} \\
\mathrm{overlap\quad matrix} &:&\mathrm{\quad}\Theta =\left(\begin{array}{ll}
\mathbf{1} & \Omega \\
\Omega ^{+} & \mathbf{1}
\end{array}
\right),\qquad \Omega _{\alpha i}=e_{\alpha}\cdot E_{i},
\label{Overlap matrix} \\
\mathrm{Hamiltonian\quad matrix} &:&\mathrm{\quad}H=\left(\begin{array}{ll}
H_{ssm \times ssm} & H_{ssm \times su3} \\
H_{su3 \times ssm} & H_{su3 \times su3}
\end{array}
\right) =\left(\begin{array}{ll}
H_{\alpha \beta} & H_{\alpha j} \\
H_{i\beta} & H_{ij}
\end{array}
\right). \label{hamiltonian Matrix}
\end{eqnarray}

In general the oblique basis results in a non-orthogonal basis that may even become linearly dependent and over-complete. Thus, the matrix equation includes an overlap matrix ($\Theta _{ij}=\langle i|j\rangle $) and has the form:
\begin{equation}
\sum_{j}\left(H_{ij}v_{j}-\lambda \Theta _{ij}v_{j}\right) =0.
\label{generalized eigenvalue problem - matrix eq.}
\end{equation}

In analogy to the SU(3) symmetry breaking in the pf-shell, which provides symmetry-adapted guidance to the relevant collective basis, one can expect that in a multi-shell space one would naturally use the symplectic symmetry group \cite{2007symplectic, 2012}. 

Furthermore, the oblique basis method can be extended and applied to the study of particle transfer reactions in strongly deformed nuclei where harmonic oscillator bound states techniques can be employed to compute the single particle binding energies in mean-field Woods-Saxon potential and then the proper tail of the wave function can be constructed using Sturmian basis that results in an equivalent oblique basis \cite{2007}.

\section{The Engineering Problems in Vibrations and Control Theory}
A system in mechanical engineering is often represented by a set of masses, springs, and dumpers \cite{2011}. The state of the system corresponds to a vector $\vec{x}$ that describes the coordinates of the masses. For our simple example we can neglect the dumper effects which usually corresponds to velocity dependent force $[c] \vec{\dot x}$. In this case the Newton's second law of motion is written as a matrix equation using mass $[m]$ and spring $[k]$ matrices: $$ [k] \vec{x}+[m] \vec{\ddot x} =0 $$ 

\subsection{Normal Modes for a Simple System}
By using the method of separation of variables and the reasonable assumption that the system will exhibit some periodic motion near its equilibrium position, one sets $\vec{x}=\vec{X} T(t)$ and  ${\ddot T}=-\omega^2 {T}$.
The resulting equation for the frequency of oscillations $\omega$ becomes effectively a generalized eigenvalue equation when the mass matrix is not diagonal. Simple example of such non-diagonal $[m]$ is a wagon with a compound pendulum inside \cite{2011}. 

Solutions $\omega_{j}$ and vectors $\vec{X_{i}}$ to the generalized eigenvalue equation: 
\begin{equation}
[k] \vec{X_{i}}-\omega_{i}^{2} [m] \vec{X_{i}} =0,
\end{equation}
form a $[k]$ or $[m]$ ``orthonormal basis'' when  $\omega_{j}\ne\omega_{i}$ ($\vec{X_{i}}^T[m]\vec{X_{j}}=0$ and $\vec{X_{i}}^T[k]\vec{X_{j}}=0$). Thus, in terms of the normal mood matrix $\hat X=\{\vec{X_{1}},\vec{X_{2}},\dots,\vec{X_{n}}\}$  the super-matrices: $M=\hat{X}^T[m] \hat{X}$ and $K= \hat{X}^T[k] \hat{X}$ become diagonal. Since the mass matrix is positive definite due to its relation to the kinetic energy of the system, then one can consider $[m]$ normalized normal modes $\vec{X_{i}}^T[m]\vec{X_{j}}=\delta_{ij}$ which results in a diagonal super-matrices $M_{ij}=\delta_{ij} \omega_{i}^2$. In terms of the normal modes the system of n-coupled masses becomes a system of  n-uncoupled masses which is equivalent to the study of n one-dimensional oscillators. Identifying the normal modes of a mechanical system is very useful once one introduces back general damping and other possible forces to the problem. Such additional forces can be treated in a perturbative like regime using the corresponding normal modes of the system.

\subsection{Evans Root-Locus Method}
How to influence the behavior of particular system to achieve the desired response is the subject of the Control Theory. In particular, by going back to our simple system but keeping only one degree of freedom we have a simple second-order differential equation. From the general theory of these equations, we know that if the system is displaced from its equilibrium there would be some oscillations and some damping that will determine the transition behavior. The response of the system to a Dirac delta input is the transfer function. Since any function can be viewed as a collection of  Dirac-delta impulses the output to a general input is essentially related to the structure of the transfer function which is in essence the ratio of the output to the input signal but in the Laplace transform variable s. Usually systems have a proportional controller K and a response of the controlled plant G(s) that is a simple polynomial function of s which result in an overall transfer function ${\cal T}(s)=K G(s)$.
	
A system is stable if all the poles of the transfer function ${\cal T}(s)$ are in the left half complex plane. A pole that has a real negative value corresponds to a pure damping without oscillations while a pole with non zero imaginary part corresponds to an oscillatory behavior. If the system has a pole with a positive real value then the system become unstable and the amplitude of a small input signal can grow to infinity. To prevent such behavior many systems include sensors $H(s)$ and negative feedback loops that reduce such output grow. Including a negative feedback modifies the transfer function of the system to become ${\cal T}(s)=K G(s)/(1+ K G(s)H(s))$.

Finding how the poles of the transfer function move in the complex plane is essential for controlling the response of a system. Often the combination of transfer functions of various control components results in a transfer function that is a ratios of polynomials. Were the controller has a simple proportional response K to the input signal. The poles of the  transfer function ${\cal T}(s)=K G(s)/(1+ K G(s)H(s))$ are clearly related to the zeros of $1+ K G(s)H(s)=0$. Understanding the zeros as K changes is thus essential. If one sets $G(s)H(s)=L(s)=\frac{b(s)}{a(s)}$ then Evans root-locus method gives rules for the location of the roots of the equation:
$$ L(s)=-\frac{1}{K}$$ as K grows from 1 to infinity. Since $L(s)$ is considered to be a proper fraction the polynomials $b(s)=\prod_{i=1}^m (s-z_i)$ and $a(s)=\prod_{i=1}^n (s-p_i)$, that is, the degree $n$ of $a(s)$ is bigger or equal to the degree $m$ of $b(s)$ then $L(s)$ has the form: 
\begin{equation}
 L(s)=\sum_{i=1}^{n}\frac{A_i}{s-p_i}=-\frac{1}{K}.
\end{equation}
This equation the main equation of the stability analysis in control theory based on Evans root-locus method \cite{1948}.

The above mathematical expression and its corresponding mathematical problem is practically the same as our Richardson-Gaudin pairing equations discussed earlier (\ref{Richeqs}). In particular, if we restrict the pairing to the problem of identical particles. Then we get the Richardson original equations \cite{1962} that only contains the pair energies $e_\alpha$ and the single particle energies $z_i=2\epsilon_i$ as some of the poles of $L$:

\begin{eqnarray}
\frac{1}{g}= \sum_{i=1}^L
\frac{1}{e_\alpha-z_i}+
\sum_{\alpha^\prime(\neq\alpha)=1}^M
\frac{2}{e_{\alpha^\prime}-e_\alpha}
\end{eqnarray}

\section{Conclusion}
In the preceding  text we have discussed the use of non-orthogonal basis sets as they emerge in the studies of complex systems like atomic nuclei and mechanical systems. In studying the nuclei, the motivation is to deal with strongly coupled competing modes where the system is far from any of the possible perturbation regimes. The well-known quadrupole-quadrupole and pairing interactions are examples of collective modes that can compete against each other as well as against the single particle spin-orbit interaction which is often considered as the main source of lifting the single particle energy degeneracy. In a sense, each of these three exactly solvable limits represents a "normal" mode of the system and can be used to address the strong coupling regime of the system. The similarity between the principle normal modes of a mechanical system and the oblique basis in nuclear physics is further enhanced by the analogy between the Richardson-Gaudin pairing equations and the Evans root-locus method from the control theory for studying the satiability of a mechanical system. The author hopes that the reader will find the analogies interesting and this work will stimulate future collaborative research on these topics and their implementations to nuclear physics and engineering problems. 

\ack
The author, V. Gueorguiev, acknowledges the generous financial support for traveling and attending the {\it Horizons of Innovative Theories, Experiments, and Supercomputing in Nuclear Physics}  (HITES 2012) conference in honor of his PhD adviser Jerry P. Draayer from Louisiana State University,  Louisiana, USA; as well as financial support provided by the Professional Development Fund for lecturers from University of California, Merced.


\section*{References}
\begin{thebibliography}{9}


\bibitem{1960} Wigner E 1960 
The Unreasonable Effectiveness of Mathematics in the Natural Sciences
{\it Comm. Pure Appl. Math.} {\bf 13} 1 (John Wiley \& Sons, New York)

\bibitem{1973} Feynman R 1973 
{\it Take the World from Another Point of View}
(Television interview, Yorkshire Great Britain)

\bibitem{2001} Gueorguiev V G, Ormand W E, Johnson C W, and Draayer J P 2002 A mixed-mode shell-model theory for nuclear structure studies {\it Phys. Rev.} C {\bf 65} 024314 ({\it Preprint} nucl-th/0110047)

\bibitem{2011} Rao S S 2011 {\it Mechanical Vibrations} (Prentice Hall)

\bibitem{1948} Evans W R 1948 Graphical Analysis of Control Systems {\it Trans. AIEE} {\bf 67} 547

\bibitem{2005} Gueorguiev V G,  Rau A R P, and Draayer J P 2006 Confined One Dimensional Harmonic Oscillator as a Two-Mode System {\it Am. J. of Phys.} {\bf 74} 394
 ({\it Preprint} math-ph/0512019)
 
 \bibitem{2007} Gueorguiev V G, Kunz P D, Escher J E, Dietrich F S 2007
Neutron Transfer Reactions for Deformed Nuclei Using Sturmian Basis
({\it Preprint} nucl-th/0706.2002)

\bibitem{2010}Gueorguiev V G 2002 
Mixed-Symmetry Shell-Model Calculations in Nuclear Physics
{\it Louisiana State University Ph.D. thesis}  ({\it Preprint} nucl-th/0704.1108)


\bibitem{1962} Richardson R W 1963 
A Restricted Class of Exact Eigenstates of the Pairing-force Hamiltonian
{\it Phys. Lett.} {\bf 3} 277; {\it Phys. Rev.} {\bf 144} 874

\bibitem{1976} Gaudin M 1976
Diagonalisation DÕune Classe DÕhamiltoniens De Spin
{\it Le Journal De Physique} {\bf 37} 1087

\bibitem{1994} Ushveridize AG 1994
{\it Quasi-Exactly Solvable Models in Quantum Mechanics}
(Institute of Physics, Bristol)

\bibitem{1997} Pan F, Draayer J P, and Ormand W E 1998
A Particle-Number-Conserving Solution to the Generalized Pairing Problem
{\it Phys. Lett.} B {\bf 422} 1 ({\it Preprint} nucl-th/9709036)

\bibitem{2002} Pan F and Draayer J P 2002 
Algebraic solutions of mean-field plus T=1 pairing interaction
{\it Phys. Rev.} C{\bf 66} 044314 

\bibitem{2002Links}
Links J, Zhou H, Gould M D, and McKenzie R H 2002
Integrability and exact spectrum of a pairing model for nucleons
{\it J. Phys.} A{\bf 35} 6459 

\bibitem{2002Asorey} Asorey M, Falceto F, and Sierra G 2002 
Chern-Simons theory and BCS superconductivity
{\it Nucl. Phys.} B{\bf 622} 593

\bibitem{2003} Pan F, Gueorguiev V G, and Draayer J P 2004
Algebraic Solutions of an Extended Pairing Model for Well-Deformed Nuclei
{\it Phys. Rev. Lett.} {\bf 92} 112503
({\it Preprint} nucl-th/0311075)

\bibitem{2006} Dukelsky J, Gueorguiev V G, Isacker P, Dimitrova S, Errea B, Lerma S 2006
Exact Solution of the Isovector Proton Neutron Pairing Hamiltonian,
{\it Phys. Rev. Lett.} {\bf 96} 072503
({\it Preprint} nucl-th/0601082)

\bibitem{2007symplectic} Dytrych T, Sviratcheva K D, Bahri C, Draayer J P, Vary J P 2007 
Evidence for Symplectic Symmetry in Ab Initio No-Core Shell Model Results for Light Nuclei
{\it Phys. Rev. Lett.} {\bf 98} 162503 ({\it Preprint} nucl-th/0704.1108)

\bibitem{2012} Draayer J P, Dytrych T, Launey K D, Langr D 2012
Symmetry-adapted no-core shell model applications for light nuclei with QCD-inspired interactions
{\it Prog. Part. Nucl. Phys.} {\bf 67} 516

\end{thebibliography}
\end{document}